\documentclass{IEEEconf}

\usepackage{graphicx}
\usepackage{color}
\newcommand{\bbibitem}{\bibitem}

\newcommand{\llabel}[1]{{\label{#1}}}

\newcommand{\ffoot}[1]{}

\renewcommand{\r}[1]{(\ref{#1})}
\newcommand{\bi}{\begin{itemize}}
\newcommand{\ei}{\end{itemize}}
\newcommand{\bd}{\begin{description}}
\newcommand{\ed}{\end{description}}
\newcommand{\be}{\begin{enumerate}}
\newcommand{\ee}{\end{enumerate}}

\renewcommand{\i}{\item}

\newcommand{\bqn}{\begin{eqnarray}}
\newcommand{\eqn}{\end{eqnarray}}
\newcommand{\eqnn}{\nonumber\end{eqnarray}}
\newcommand{\eqnl}[1]{\llabel{#1}\end{eqnarray}}

\newcommand{\nn}{\nonumber}

\newcommand{\ba}[1]{\begin{array}{#1}}
\newcommand{\ea}{\end{array}}

\font\tenmsb=msbm10
\font\sevenmsb=msbm7
\font\fivemsb=msbm5
\newfam\msbfam
\textfont\msbfam=\tenmsb
\scriptfont\msbfam=\sevenmsb
\scriptscriptfont\msbfam=\fivemsb
\def\Bbb#1{{\fam\msbfam\relax#1}}

\newcommand{\R}{\Bbb{R}}
\newcommand{\C}{\Bbb{C}}

\newcommand{\fine}{\end{document}}

\def \trait (#1) (#2) (#3){\vrule width #1pt height #2pt depth #3pt}
\def \qed{\hfill
        \trait (0.1) (6) (0)
        \trait (6) (0.1) (0)
        \kern-6pt   
        \trait (6) (6) (-5.9)
        \trait (0.1) (6) (0)
\medskip}

\newtheorem{ml}{\bf Lemma}
\newtheorem{Theorem}{\bf Theorem}

\newtheorem{mrem}{\bf \underline{{\sl Remark}}}
\newtheorem{mcc}{\bf Corollary}
\newtheorem{Definition}{\bf Definition}
\newtheorem{mpr}{\bf Proposition}
\newtheorem{mproperty}{\bf Property}

\newcommand{\bt}{\begin{Theorem}}
\newcommand{\et}{\end{Theorem}}
\newcommand{\bl}{\begin{ml}}
\newcommand{\el}{\end{ml}}
\newcommand{\bp}{\begin{mpr}}
\newcommand{\ep}{\end{mpr}}
\newcommand{\bc}{\begin{mcc}}

\newcommand{\bproperty}{\begin{mproperty}}
\newcommand{\eproperty}{\end{mproperty}}

\newcommand{\ec}{\end{mcc}}
\newcommand{\bdeff}{\begin{Definition}}
\newcommand{\edeff}{\end{Definition}}
\newcommand{\brem}{\begin{mrem}\rm}
\newcommand{\erem}{\end{mrem}}

\newcommand{\ppotR}[3]
{

\begin{figure}\begin{center}
~\includegraphics[width=#3truecm]{./#1.eps}\\
\caption{#2}
\llabel{#1}
\end{center}
\end{figure}
\noindent$\!\!$}
       
\newcommand{\al}{\alpha}

\newcommand{\om}{\omega}

\renewcommand{\th}{\theta}

\newcommand{\sceq}{Schr\"{o}dinger equation\ }

\newcommand{\und}{\underline}

\newcommand{\vep}{\varepsilon}

\newcommand{\vect}[3]{\left(\begin{array}{ccc}{#1}\\{#2}\\{#3}\end{array}\right)}

\begin{document}
\pagestyle{empty}

\thispagestyle{empty}
\title{Time Minimal Trajectories for two-level Quantum Systems
with Drift}
\author{  
Ugo Boscain, Paolo Mason\\
{\footnotesize SISSA, via Beirut 2-4 34014 Trieste, Italy}\\
e-mails {\tt boscain@sissa.it, mason@sissa.it}
}
\maketitle
\begin{abstract}

On a two-level
quantum system driven by an external field, we consider the population 
transfer problem from the first to the second level,
minimizing the time of transfer, with bounded field amplitude.
On the Bloch sphere (i.e. after a suitable Hopf projection), this problem
can be attacked with techniques of optimal syntheses on 2-D manifolds.

Let $(-E,E)$ be the two energy levels, and $|\Omega(t)|\leq M$ the bound on the
field amplitude.
For each values of $E$ and $M$,  we provide the explicit
expression of the time optimal trajectory steering the state one to the
state two in terms of a parameter that should be computed numerically.

For $M<<E$, every time optimal trajectory is 
periodic (and in particular bang-bang) with frequency of the order of the 
resonance frequency $\om_R=2E$.

On the other side, for $M>E$
the time optimal trajectory steering the state one to the state two is 
bang-bang with exactly one switching. Fixed $E$ we also prove that for 
$M\to\infty$ the time needed to reach the state two tends to zero. 

Finally we compare these results with some known results 
of Khaneja, Brockett and
Glaser and with those  obtained in the 
Rotating Wave Approximation. 

\ffoot{
On the other side, for $M>E$
every time optimal trajectory is of
the type bang-singular-bang, where the singular trajectory is a trajectory
of the drift (the Hamiltonian with no external fields), and, in a sense,
we recover, as a particular case, a known results of Khaneja, Brockett and
Glaser.
}

\end{abstract}


\section{Introduction}

In this paper we apply techniques
of optimal synthesis on 2-D manifolds to
the population
transfer problem in a two-level
quantum 
system  (e.g. a spin 1/2 particle) driven by an external field (e.g. a 
magnetic field along a fixed axis).
Two-level systems are the simplest quantum mechanical models  interesting 
for applications (see for instance \cite{allen,choen}).
The dynamics is governed by the time dependent Schr\"odinger equation (in
a system of units such that $\hbar=1$):
\begin{eqnarray}
i\frac{d\psi(t)}{dt}=H(t)\psi(t),
\label{se}
\end{eqnarray}
where $\psi(.)=(\psi_1(.),\psi_2(.))^T:[0,T]\to{{\C}}^2$,
$\sum_{j=1}^2|\psi_j(t)|^2=1$ (i.e. $\psi(t)$ belongs to the sphere
$S^3\subset {{\C}}^2$), and:
\begin{eqnarray}
H(t)=\left(\begin{array}{cc} -E& \Omega(t) \\ \Omega(t)
&E
\end{array}\right),
\label{hg1}
\end{eqnarray}
where $E$, is a real number ($\pm E$ represent the {energy levels} of the 
system). The \underline{control}
$\Omega(.)$,  that we assume  to be a real function, different
from zero only in a fixed interval, represents the
{external pulsed field}. In the following we call \und{drift 
term}, the  Hamiltonian with no 
external fields (i.e., the term $diag(-E,E)$).

The aim
is to induce a transition from the first level (i.e., $|\psi_1|^2=1$) to 
the second level (i.e., $|\psi_2|^2=1$), minimizing
the time of transfer, with bounded field amplitude:
\begin{eqnarray}
|\Omega(t)|\leq M,~~~\mbox{for every $t\in[0,T]$},
\label{bound}
\end{eqnarray}
where $T$ is the time of the transition and $M$ is a positive real
constant
representing the maximum amplitude
available.

\brem
{\sl This problem was studied also in \cite{daless}, but with
quadratic cost $\int_0^T \Omega(t)^2~dt$ and with no bound on the control.
In this case, optimal solutions 
can be expressed in terms of Elliptic functions.}
\erem

It is a standard fact to eliminate an irrelevant global factor of phase by 
projecting the system on a two dimensional real sphere $S^2$ (called the 
Bloch Sphere) by means of an Hopf map. In this way the \sceq\ \r{se}, 
\r{hg1} becomes the single input affine system (after setting 
$u(t)=\Omega(t)/M$):
\begin{eqnarray}
\dot{y}&=&F_S(y)+uG_S(y),\mbox{ ~~where:}\llabel{cs-p}\\
&&y=(y_1,y_2,y_3)\in\R^3,~~~\sum_{j=1}^3 y_j^2=1\\
&&|u|\leq1, \llabel{cs-bound}\\
&&F_S(y):=k \cos(\al)\left(\begin{array}{c} 
-y_2\\y_1\\0\end{array}\right),               \label{cs-F} \\
&&G_S(y):=k\sin(\al)\left(\begin{array}{c} 
0\\-y_3\\y_2\end{array}\right),    \label{cs-G}\\
&&\alpha=\arctan \left( \frac{M}{E}\right)\in\,]0,\pi/2[,\llabel{cs-u}
\end{eqnarray}
and the constant $k$ is given by $k=2E/\cos(\al)=2 \sqrt{M^2+E^2}$.

{\bf Normalizations} In the following, to simplify the notations, we 
normalize $k=1$. This normalization corresponds to a reparametrization of the 
time. More precisely, if $T$ is the minimum time to steer the state one to 
the state two for the system with $k=1$, the corresponding minimum time for 
the original system is simply $T/\sqrt{M^2+E^2}$. We come back to the original 
value of $k$ only in Section \ref{s-comparison}.\\\\
The vector fields $F_S(y)$ and $G_S(y)$ describe rotations respectively
around the axes $y_3$ and $y_1$.
Now the state one is represented by the point $y_3=1$ (called the 
\und{north pole}) and the state two  by the point $y_3=-1$ 
(called the \und{south pole}). The optimal control problem is then  to 
connect the north pole to
the south pole in minimum time. As usual we assume the control $u(.)$ to 
be a measurable function satisfying \r{cs-bound} almost everywhere. 
The  corresponding trajectory is a Lipshitz continuous 
function $y(.)$ satisfying \r{cs-p} almost everywhere. 

\bigskip
The most important and powerful tool to study  optimal 
trajectories 
is the well known Pontryagin Maximum Principle (in the following PMP, see 
for instance \cite{agra-book,pontlibro}). It is a first order 
necessary condition for optimality and generalizes the Weierstra\ss \ 
conditions of Calculus of Variations to problems with non-holonomic 
constraints. For each optimal trajectory, the PMP 
provides a lift to the cotangent bundle that is a solution to a suitable
pseudo--Hamiltonian system.\\
Even if the PMP is powerful, giving a complete solution to an  
optimization problem remains extremely difficult. 
First, one is faced with the 
problem of integrating a Hamiltonian system. 
Second, one should manage 
with ``non Hamiltonian solutions'' of the PMP, the so called 
\underline{abnormal 
extremals}. Finally,  even if one is able to find all the solutions of the 
PMP it remains the problem of {selecting}, among them, the
{optimal trajectories}.
For these reasons, usually, one can hope to find  a complete solution of 
an optimal control problem   for very special 
costs, dynamics and in low dimension only.

Two dimensional minimum time problems in control affine form, (like the 
problem \r{cs-p}--\r{cs-u}) are nice cases for which 
the analysis can be pushed much further, thanks to the theory developed in 
\cite{libro} (see also the references therein).
In this paper we take advantage of that theory to 
restrict the set of candidate optimal trajectories.
The optimal trajectories are then identified, by requiring that they 
respect certain crucial symmetries of the system.
More precisely, for $M>E$ the time optimal trajectories steering the state one 
to the state two are bang-bang with exactly one switching, and we give the 
exact expressions of the corresponding optimal controls. In particular, fixed
$E$ we see that for $M\to\infty$ the time needed to reach the state two tends 
to zero.\\ 
On the other side, for $M<<E$, every time optimal trajectory is 
periodic (and in particular bang-bang) with frequency of the order of the 
resonance frequency $\om_R=2E$, and can be selected among a finite set 
of trajectories which corresponds to solutions of suitable equations.

\ffoot{parlare del fatto che la soluzione e' un problema globale, ma noi 
usando le simmetrie non abbiamo bisogno di risolvere tutto.
{\tt dire che la costruzione della sintesi, sebbene possibile e' molto 
difficile. Dire che una costruzione parziale e' stata fatta per $\al$ 
piccolo. Dire che per $\al$ grande sara' presentata in 
Dire che 
qui ci interessiamo solo alla traiettoria che arriva al punto finale.}
}

\brem
\label{r-rwa}
If we were describing 
either a spin 1/2 particle  driven by 
two magnetic fields (one along the $x$ axis and one along the $y$ axis) or 
a two-level molecula driven by an external field in the Rotating Wave 
Approximation (RWA for short, see for instance \cite{allen}), then our 
Hamiltonian would contain  complex controls:
\begin{eqnarray}
H(t)=\left(\begin{array}{cc} -E& \Omega(t) \\ \Omega^\ast(t)
&E
\end{array}\right),
\label{hgcomplex}
\end{eqnarray}
where  $(^\ast)$ indicates the complex conjugation involution.
In this case the minimum time problem with bounded controls (i.e., 
$|\Omega(t)|\leq M$)  is easier,\ffoot{vedere se dire che 
e' equivalente all'energia} since it is possible to 
eliminate the drift term  by a unitary change of coordinates and a change
of controls (interaction picture). This problem has been studied in 
\cite{q1,daless}. The simplest time optimal trajectory, steering the 
system from the state one to the state two, corresponds to controls in 
\und{resonance} with the energy gap $2E$, and with maximal amplitude i.e.
\bqn
\Omega(t)=Me^{i(2E)t}
\eqn
The quantity $\om_R=2E$ is called the \underline{resonance frequency}. 
In this case, the time $T_{RWA}$ of transfer is proportional to the 
inverse 
of the laser amplitude. More precisely $T_{RWA}=\pi/(2 M)$, see for 
instance 
\cite{q1}. In Section \r{s-comparison} the minimum time of 
transfer for the  
Hamiltonians \r{hg1} and \r{hgcomplex} are compared.
\erem

\medskip
In Section \ref{s-known}, we recall some basic properties of optimal 
trajectories for the system \r{cs-p}--\r{cs-u}, that were already 
obtained in \cite{dubin-yac}.
In Section \ref{s-main} we state our main results and in \ref{s-comparison} we
compare these results with some known results of Khaneja, Brockett and
Glaser and with those  obtained in the Rotating Wave Approximation.
In Section \ref{s-sketch} we give an idea of the techniques we used.
\section{Known Results}\llabel{s-known}
Recall that we have normalized $k=1$.
Note that, since the system \r{cs-p}--\r{cs-u} is Lie bracket generated on 
a compact 
manifold and the set of velocities  is compact and convex  then, for each 
pair of points $p$ and $q$ belonging to 
$S^2$, there exists a time optimal trajectory joining $p$ to $q$.

Although, with different 
purposes, The minimum time problem for the control system \r{cs-p}, 
\r{cs-u}, although with different purposes, has been partially studied in 
\cite{dubin-yac} (see also \cite{dubin-cdc}). 

In particular in \cite{dubin-yac} it was proved that 
every time optimal trajectory is a finite concatenation of bang arcs 
(i.e., corresponding to  control a.e. constantly equal to $+1$ or $-1$)  
and singular arcs (i.e., corresponding to singularities of the End point 
mapping, see for instance \cite{bon}, that in our case correspond to
controls a.e. vanishing) with 
some special structure.  More precisely:
\bdeff
{\sl A control
$u:[a,b]\to[-1,1]$ is said to be \und{bang-bang} if $u(t)\in \{-1,1\}$
a.e. in  $[a,b]$. Moreover, if $u(t)\in \{-1,1\}$ and $u(t)$ is constant
for almost
every  $t\in[a,b]$, then $u$ is called a \und{bang} control. 
If $u(t)=0$ for almost
every  $t\in[a,b]$, then $u$ is called a \und{singular} control.
 
A \und{switching} time of $u$ is a time $t\in[a,b]$ such that,  
for every $\vep>0$, $u$ is not
bang on $(t-\vep,t+\vep)\cap [a,b] $. A control with a finite 
number of switchings
is called \und{regular bang-bang}. A trajectory of the control system
\r{cs-p}--\r{cs-u} is a bang
trajectory, singular trajectory,  
bang-bang trajectory, regular bang-bang trajectory respectively, if it
corresponds to a bang control, singular control, bang-bang control, 
regular bang-bang 
control respectively.}\ffoot{vedere se non si piu' eliminare un po' di 
roba che non serve} 
\edeff

In the sequel, we use the following convention. The letter $B$
refers to a bang arc and the letter $S$ refers to a singular
arc. A concatenation of bang and singular arcs
is labeled by the corresponding letter sequence, written in order 
from left to right.
Sometimes, we will use a subscript to indicate the time duration of an 
arc,
so that we use $B_t$ to refer to a bang arc defined on an interval
of length $t$ and, similarly, $S_t$ for a singular arc defined on
an interval of length $t$.

\bigskip
Using the PMP and the Theory developed in \cite{libro}, in 
\cite{dubin-yac} (see also \cite{agra-sympl-x})  it was
proved the following:
\bp
{\sl Consider the control system \r{cs-p}--\r{cs-u}. Then: 
\bd
\i[A.] every time optimal trajectory is a finite concatenation of bang and 
singular arc; 

\i[B.] if  a time optimal trajectory contains a  bang arc $B_t$, then 
   $t<2\pi$;

\i[C.] if a time optimal trajectory contains a singular arc, then   it is 
of the type $B_tS_sB_{t'}$, with $s\leq\frac{\pi}{\cos(\al)}$, $t,t'\geq 0$. 
Moreover the support of singular arcs lies on the set (called equator) 
$y_3=0$.

\i[D.] if  a time optimal trajectory is bang-bang, then  
the time duration $\bar T$ along an interior bang arc is the same for all 
interior bang arcs and verifies
$\pi\leq\bar T< 2\pi$.
\ffoot{
vedere poi se dire che la singolare non puo' passare attraverso un punto 
in cui $G=0$. Vedere se dare i segni dei possibili switchings nelle 
diverse zone.
}
\ed}
\ep
From {\bf C.} it follows that the first and the last arc on optimal   
trajectory connecting the north with the south pole are not singular.
The following proposition (see \cite{dubin-yac} for the proof) gives more 
details on the optimal trajectories  starting at the north pole in the 
case $\al<\pi/4$.

\bp\llabel{p-sbang}
{\sl Consider the control system \r{cs-p}--\r{cs-u}, and assume 
$\al<\pi/4$. Then the optimal 
trajectories starting from the north pole 
are of the form 
$B_{s_i}B_{v({s_i})}\cdots B_{v({s_i})}B_{s_f}$, where $s_i\in[0,\pi]$,
$s_f\in[0,v(s_i)]$ and
\begin{equation}\label{vs}
v(s_i)=\pi+2 \arctan\left(\frac{\sin(s_i)}{\cos (s_i)+\cot^2(\alpha)}\right).
\end{equation}
}
\ep
Note that the function $v(.)$ is such that $v(0)=v(\pi)=\pi$ and moreover
it is increasing on the interval $[0,\bar{t}]$ and decreasing on 
$[\bar{t},\pi]\,$, where $\bar{t}=\arccos(-\tan^2(\al))$, moreover
if $\al$ is small the maximum of $v(.)$ is $v(\bar{t})=2\arccos(-\tan^2(\al))
\sim\pi+2\al^2$ 
(see Figure~\ref{vds}). \\
\begin{center} 
\begin{figure}
\input{vudiesse-comp.pstex_t}
\caption{Graph of $v(.)$ when $\al=\pi/6$}
\llabel{vds}
\end{figure}
\end{center}

\ffoot{dire che per $\al\geq\pi/4$,  
una traiettoria che parte dal polo nord e arriva al polo sud, non puo' 
contenere un arco singolare. Questo forse bisognerebbe dimostrarlo.
}

\section{Main Results}\llabel{s-main}
\subsection{The $\al\geq\pi/4$ Case}
In the case  $\al\geq\pi/4$, there are exactly four optimal trajectories 
steering the state one to the state two. They are easily described by the 
following:
\medskip
\bp
\label{propA}
{\sl Consider the control system \r{cs-p}--\r{cs-u}, and assume
$\al\geq\pi/4$. Then the optimal trajectories steering the north pole to the 
south pole are bang-bang with only one switching. More precisely they are 
the four trajectories corresponding to the four controls
\bqn
u^{(1)}=\left\{
\ba{l} 
u=1,~t\in[0,s_A]\\
u=-1,~t\in]s_A,T]\\
\ea
\right.,~~u^{(2)}=\left\{
\ba{l} 
1,~t\in[0,s_B]\\
-1,~t\in]s_B,T]\\
\ea
\right.\nn\\
u^{(3)}=\left\{
\ba{l} 
-1,~t\in[0,s_A]\\
1,~t\in]s_A,T]\\
\ea
\right.,~~u^{(4)}=\left\{
\ba{l} 
-1,~t\in[0,s_B]\\
1,~t\in]s_B,T]\\
\ea
\right.\nn
\eqn
where: 
\bqn
&&s_A=\pi
-\arccos(\cot^2(\al)),\nn\\
&&s_B=\pi
+\arccos(\cot^2(\al)),\nn\\
&&T=2\pi.\llabel{T1}
\eqn
}
\ep
\medskip
One can easily check that the switchings described in Proposition  
\ref{propA} occours on the equator ($y_3=0$). Figure \ref{f-propA} 
represents the optimal controls $u^{(1)}$ and 
$u^{(2)}$.
\begin{center} 
\begin{figure}
\input{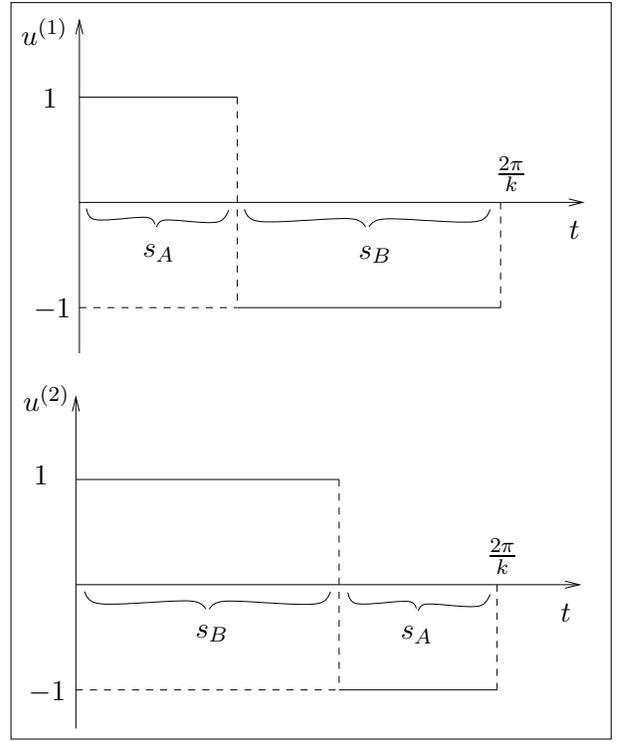}
\caption{Case $\al>\pi/4$: optimal trajectories}
\llabel{f-propA}
\end{figure}
\end{center}
\subsection{The $\al<\pi/4$ Case}
If $\al<\pi/4$, the situation is more complicated. From Proposition 
\ref{p-sbang}, we know that every optimal trajectory starting at 
the north pole has the form:
\bqn
B_{s_i}\underbrace{B_{v({s_i})}\cdots B_{v({s_i})}}_{n-1\ times  
}B_{s_f}
\eqn
where the function $v(s_i)$ is given by formula \r{vs}. (In the following 
we do not specify if the first bang corresponds to control $+1$ or $-1$, 
since, as a consequence of the symmetries of the problem, if $u(t)$ is 
an optimal control steering the north pole to the south pole, $-u(t)$ 
steers the north pole to the south pole as well.)  

It remains to identify one or more values of $s_i,s_f$ and the 
corresponding  number of switching $n$ for this trajectory to reach the 
south pole.

Next, given $s\in[0,\pi]$ such that $s\neq\bar{t}=\arccos(-\tan^2(\al))$ we 
call $s'(s)$ the unique solution to the equation $v(s)=v(s'(s))$ with 
$s'(s)\neq s$ and we define $s'(\bar{t})=\bar{t}$ (see also Figure 
\ref{vds}). 
Considering the symmetries of the problem, one can prove 
that if $\al<\pi/4$,  $s_f$ is equal either to $s_i$ or to $s'(s_i)$.
This fact  is described by Lemma \ref{v=v} below.

In the following we describe how to identify candidate 
triples $(s_i,s_f,n)$ for which the corresponding trajectory steers the 
north pole to the south pole in minimum time.
There are two kind of candidate optimal trajectories. 
\bi
\i $s_f=s'(s_i)$, called TYPE-1-candidate optimal trajectories
\i $s_f=s_i$ called TYPE-2-candidate optimal trajectories
\ei
\bigskip
Define the following functions:
\bqn
\th(s)=2 \arccos\!\left(\sin^2\!\left(\frac{v(s)}{2}\right)
\cos (2\al)-\cos^2\!\left(\frac{v(s)}{2}\right)\right)
\eqnl{thth}
\bqn
\beta(s)=2 \arccos (\sin(\al)\cos(\al) (1-\cos(s)))
\eqnl{bebe}
\bp \llabel{p-1}{\bf (TYPE-1-trajectories)}
{\sl Let $\al<\pi/4$ and $s\in[0,\pi]$. Fixed $\al$, the following 
equation for 
the 
couple $(s,n)$:
\bqn
\mathcal{F}(s):=\frac{2\pi}{\th(s)}=n,
\eqnl{effe}
has either two or zero solutions. More precisely if $(s,n)$ is a 
solution to equation \r{effe}, then $(s'(s),n)$ is the second one, and   
the trajectories $B_{s}\underbrace{B_{v({s})}\cdots 
B_{v({s})}}_{n-1}B_{s'(s)}$ and $B_{s'(s)}\underbrace{B_{v({s})}\cdots
B_{v({s})}}_{n-1}B_{s}$  are 
TYPE-1-candidate 
optimal trajectories.} 
\ep

In figure \ref{disegnoF} the graph of the function \r{effe} is drawn for 
a particular value of $\al$, namely $\al=0.13$.
\ppotR{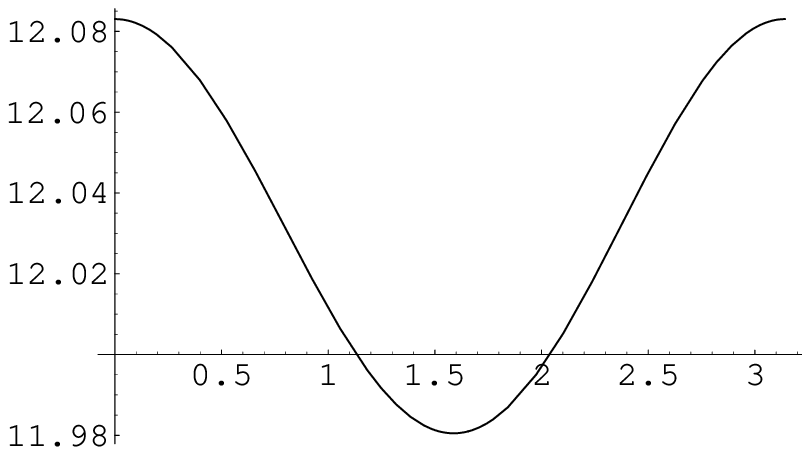}{Graph of the function $\mathcal{F}$ with $\al=0.13$}{6}
\bp \llabel{p-2} {\bf (TYPE-2-trajectories)}
{\sl Let $\al<\pi/4$ and $s\in[0,\pi]$. Fixed $\al$, the following 
equation for 
the 
couple 
$(s,n)$:
\bqn
\mathcal{G}(s):=\frac{2\beta(s)}{\th(s)}+1=n,
\eqnl{gi}
has exactly two solutions. More precisely these solutions have the form 
$(s_1,n)$, $(s_2,n+1)$ and the trajectories 
$B_{s_1}\underbrace{B_{v({s_1})}\cdots B_{v({s_1})}}_{n-1}B_{s_1}$ and  
$B_{s_2}\underbrace{B_{v({s_2})}\cdots B_{v({s_2})}}_{n}B_{s_2}$ 
are TYPE-2-candidate
optimal trajectories.}
\ep

In figure \ref{disegnoG} the graph of the function \r{gi} is drawn for 
$\al=0.13$.
\ppotR{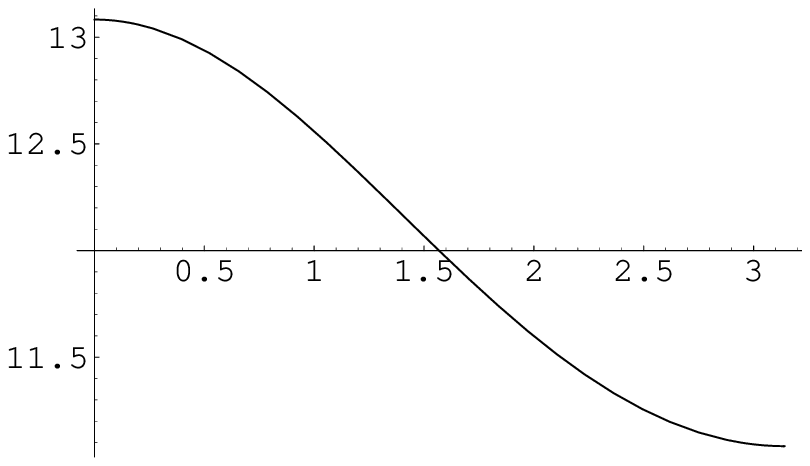}{Graph of the function $\mathcal{G}$ with $\al=0.13$}{6}

Propositions \ref{p-1} and \ref{p-2} select a set of (possibly coinciding) 
$4$ or $8$ candidates  optimal trajectories (half of them starting with control $+1$ and the 
other half with control $-1$) corresponding to triples $(s_i,s_f,n)$ that 
can be easily computed numerically.
 
Then the optimal trajectories can be easily selected. Notice that there 
are at least two optimal trajectories steering the north to the south 
pole (one starting with control $+1$ and the other with control $-1$).

In the particular case in which $\pi/(2\al)$ is an integer number $\bar{n}$ 
one can see that TYPE-1 candidate optimal trajectories coincide with  some 
of TYPE-2 candidate optimal trajectories. They are of the 
type $B_\pi \underbrace{B_\pi ...B_\pi}_{\bar{n}-2} B_\pi$ or of the type
$B_s \underbrace{B_{v(s)} ...B_{v(s)}}_{\bar{n}-1} B_s$ for some $s\in]0,\pi[$.

Otherwise if $\pi/(2\al)$ is not an integer number, define:
$$
m:=[\frac{\pi}{2\al}],~~~~r:=\frac{\pi}{2\al}-m\in[0,1[
$$
where $[.]$ denotes the integer part. One can prove the following:
\bp
There exists $\bar r(m)\in]0,1[$ such that:
\bi
\i if $r\in[0,\bar r(m)]$ then equation \r{effe} admits exactly two 
solutions  that are both optimal, while TYPE-2 candidate optimal 
trajectories are not.
\i if $r\in]\bar r(m),1[$ then equation \r{effe} does not admit any 
solution. 
\ei 
\ep
The claims on existence of solutions of the previous propositions come 
from 
the fact that  $\mathcal{F}(0)=\mathcal{F}(\pi)=\frac{\pi}{2 \al}$ and the 
only minimum point of $\mathcal{F}$ occurs at $\bar{s}=\pi-\arccos(\tan^2(\al))$.\\ 
It turns out that the image of $\mathcal{F}$ is a small interval whose
length is of order $\al^3$ and therefore equation \r{effe} has a 
solution only if $\al$ is close enough to $\frac{\pi}{2n}$ for some integer 
number $n$.\\
On the other hand it is possible to estimate the derivative of $\mathcal{G}$ 
with respect to $s$ showing that it is negative in the open interval 
$]0,\pi[$.
Therefore, since $\mathcal{G}(0)=\frac{\pi}{2 \al}+1$ and 
$\mathcal{G}(\pi)=\frac{\pi}{2 \al}-1$, equation \r{gi} has always two 
solutions (if $\frac{\pi}{2 \al}$ is an integer number then the trajectories 
corresponding to the solutions $s_i=0$ and $s_i=\pi$ coincide).\\

Using the previous analysis one can easily prove the following:
\bp
{\sl
If $N$ is the number of switchings of an optimal trajectory joining the 
north to the south pole, then
$$\frac{\pi}{2\al}-1\leq N<\frac{\pi}{2\al}+1.  $$}
\ep

Using these inequalities and the fact that the function
$\ \displaystyle{2s+\left(\frac{\pi}{2\al}-1\right)v(s)}\ $ is increasing on 
$[0,\pi]$, we can give a rough estimate of the time 
needed to reach the south pole:

\bp
{\sl The total time $T$ of an optimal trajectory joining the north to the 
south 
pole satisfies the inequalities:
\bqn
\frac{\pi^2}{2\al}-2\pi<T<\frac{\pi^2}{2\al}+\pi. 
\eqnl{T2}}
\ep
\subsection{Comparison with results in the RWA and with \cite{brokko}}
\llabel{s-comparison}
In this section we come back to the original value of $k$ i.e. 
$k=2E/\cos(\al)=2\sqrt{M^2+E^2}$, and 
we compare the time necessary to steer the state one to 
the state two for our model and  the model (in the RWA) described 
in Remark \ref{r-rwa}. 

For our model we have the following:
\bi
\i for $\al\geq\pi/4$ then $T=2\pi/k=\pi/\sqrt{M^2+E^2}$;
\i for $\al<\pi/4$  then $T$ is estimated by 
$$
\frac1k\left(\frac{\pi^2}{2\al}-2\pi\right)<T<\frac1k
\left(\frac{\pi^2}{2\al}+\pi\right).
$$
\ei
On the other hand, for the model in the RWA, we have 
$T_{RWA}=\pi/(2M)$  (cfr. Remark \ref{r-rwa}). 

Fixed $E=1$, in Figure \ref{f-innominata} the times $T$ and 
$T_{RWA}$ as function of $M$ are compared. Notice that although $T_{RWA}$ is bigger than the lower 
estimate of $T$  in some interval, we always have $T_{RWA}\leq T$. This is 
due to the fact that the admissible velocities of our model are a subset 
of the admissible velocities 
of the model in the RWA. 

Notice that, fixed $E=1$, for $M\to 0$ we have  $T\sim \pi^2/(4 M)$, while 
for $M\to \infty$, we have   $T\sim \pi/M$. In other words:
\bi
\i for  $M\to 0$ we have $T\sim (\pi/2) T_{RWA}$,
\i for  $M\to \infty$ we have $T\sim 2 T_{RWA}$.
\ei

\brem
For $M<<E$ (i.e. for $\al$ small) $v(s)\sim \pi/(2E)$. It follows that a 
time optimal trajectory connecting the north to the south pole 
(in the interval between the first and the last bang)
is periodic with period $P\sim\pi/E$ i.e. with a frequency of the order 
of the resonance frequency $\om_R=2E$ (see Figure \ref{f-confr-opt}).
On the other side if $M>E$ then the time optimal trajectory connecting the 
north with the south pole is the concatenation of two pulses. Notice that 
if $M>>E$, the time of transfer is of the order of $\pi/M$ and 
therefore tends to 
zero as $M\to\infty$. It is interesting to compare this result with a  
result of Khaneja, Brockett and Glaser, for a two level system, but 
with no bound on controls (see \cite{brokko}).
They estimate the infimum time to reach every point of whole group $SU(2)$ 
in $\pi/E$.

Indeed for our model it is possible to prove that, for $M\to\infty$,
not every point of the Bloch sphere can
be reached from the state one in an
arbitrarily small time, but this is the case for the
state two, as we discussed above.
\erem


\begin{figure}
\input{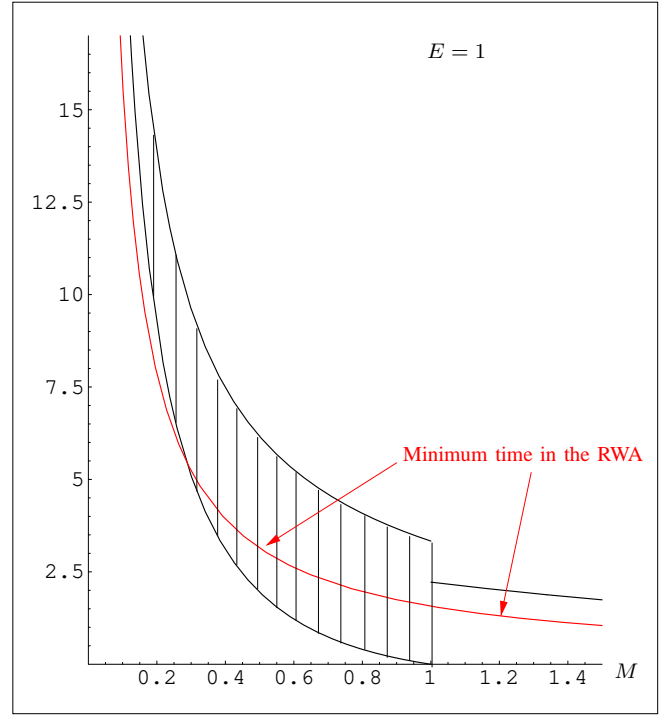}
\caption{Estimate on the minimum time to reach the state two and 
comparison with the time needed in the RWA}
\llabel{f-innominata}
\end{figure}

\begin{figure}
\input{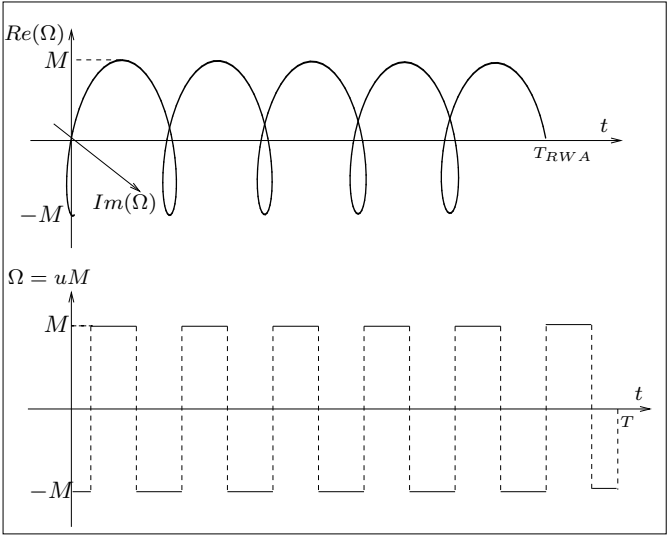}
\caption{Comparison between the optimal strategy for our system and in the RWA}
\llabel{f-confr-opt}
\end{figure}

\section{Sketch of the Proofs}\llabel{s-sketch}
In this section we give an idea of the techniques we used to prove our 
results. The key point is described by the following Lemma which states a 
property of optimal trajectories as a consequence of the symmetries of the 
problem. Recall that we have normalized $k=1$.
Before stating the Lemma we note that it is possible to extend Proposition 
\ref{p-sbang} to the case $\al\geq\pi/4$, assuming that
$s_i\in[0,\arccos(-\cot^2(\al))[$ , and 
in this case $v(.)$ is an increasing function on its interval of definition. 
\medskip
\bl
{\sl Every optimal bang-bang trajectory, connecting the north to the south 
pole,  with more than one switching is such
that $v(s_i)=v(s_f)$ where $s_i$ is the first switching time and $s_f$ is the 
time needed to steer the last switching point to the south 
pole.\label{v=v}}
\el
\medskip
{\bf Proof of the lemma.} Consider the problem of connecting the south pole
to the north pole in minimum time through the system
\bqn
\dot{z}=F'_S(z)+uG'_S(z)
\eqnl{rev}
where $z\in S^2 $ and $F'_S(z)=-F_S(z)$, $G'_S(z)=-G_S(z)$.\\
The trajectories of system \r{rev} coincide with those of the 
system \r{cs-p}--\r{cs-u}, but the velocity is reversed. Therefore the optimal 
trajectories for the new problem coincide with the optimal ones for the
system \r{cs-p}--\r{cs-u} connecting the north pole to the south pole, and the 
time between two switchings is the same.\\
Moreover, if we perform the change of coordinates $(z_1,z_2,z_3)\rightarrow 
(y_1,y_2,y_3)=(-z_1,z_2,-z_3)$, then the new problem becomes exactly the 
starting problem, and so we deduce that, if we have more than one switching, 
it must be $v(s_{i})=v(s_{f})$.

\medskip
{\bf Proof of Proposition \ref{propA}.} First one can easily see 
that the only possible trajectories steering the north to the south 
pole, with only one 
switching are those described by the proposition. So, since by Proposition
\ref{p-sbang}
the total time for trajectories with more than two switchings is larger than 
$2\pi$, it remains to compare our candidate with the
trajectories containing a singular arc and with those with exactly two 
switchings. In the first case the trajectories must be of the type 
$B_tS_sB_{t'}$ and the only possibility is 
$t=t'=\pi-\arccos(\cot^2(\al))$, while the total time is
$\, 2\pi-2\arccos(\cot^2(\al))+2\arccos(\cot(\al))/\cos(\al)$ 
which is larger than $2\pi$.\\
One can observe that $v(.)$ is an increasing function if $\al>\pi/4$ and 
therefore, if we apply Lemma \ref{v=v}, we obtain that for an optimal 
trajectory with more than one switching it must be $s_i=s_f$. In particular 
the bang-bang trajectories with exactly two switchings joining the north pole 
to the south pole and with $s_i=s_f<\pi$ can be explicitly determined and
their corresponding total time is $\,2\pi+2\arcsin
\left(\frac{1}{2\sin(\al)}\right)$.

\medskip
{\bf Proof of Propositions \ref{p-1} and \ref{p-2}.} If $\al<\pi/4$ then 
$v(0)=v(\pi)=\pi$,
moreover $v(.)$ is increasing between $0$ and $\arccos (-\tan^2 \al)$ and 
decreasing between $\arccos (-\tan^2 \al)$ and $\pi$ (see Figure \ref{vds}).\\
Therefore, given $\al<\pi/4$ and $s\in [0,\pi]$ with $s\neq\arccos (-\tan^2 
\al)$, there exists one and only one time $s'(s)\in[0,\pi]$ different from 
$s$, such that $v(s)=v(s'(s))$.\\ 
Notice that $s$ and $s'(s)$ satisfy the nice property
\bqn 
s+s'(s)= v(s). 
\eqnl{belin}
Indeed both $s$ and $s'(s)$ satisfy the following equation in $t\in[0,\pi]$:
$$
\cot\left(\frac{1}{2}v(s)\right)=-\frac{\sin (t)}{\cos 
(t)+\cot^2\al}\Rightarrow$$
$$
\Rightarrow\quad \cos\left(\frac{1}{2}v(s)-t\right)=-\cos
\left(\frac{1}{2}v(s)\right)\cot^2\al .
$$
Therefore, since 
$\,\ \frac{1}{2}v(s)-t\in[-\pi ,\pi ]\quad\forall s,t\in[0,\pi]\,$ and 
$s'(s)\neq s$, it must be: 
$$s'(s)-\frac{1}{2}v(s)=\frac{1}{2}v(s)-s\quad\Rightarrow\quad
s+s'(s)=v(s).$$
So we deduce that
there are two possible cases: 
\be
\i[($\clubsuit$)] $s_f=s'(s_i)$
\i[($\spadesuit$)] $s_f=s_i$
\ee

The description of candidate
optimal trajectories is simplified by 
the following Lemma, of which we skip the proof.
\bl
Set: $$Z(s)=\frac{1}{\rho}
\left(\begin{array}{ccc}
0 & \cot\left(\frac{1}{2}v(s)\right) & -\sin(\al) \\
-\cot\left(\frac{1}{2}v(s)\right) & 0 & 0  \\
\sin(\al)  & 0 & 0 
\end{array}\right)
$$
$$\mbox{where}\quad\quad\rho=\sqrt{\cot^2\left(\frac{1}{2}v(s)\right)+
\sin^2(\al)}$$
Then, if $\th(s)$ is defined as in \r{thth}, and $X^+:=F_S+G_S$, $X^-:=F_S-G_S$
, we have: $$e^{\th (s) Z(s)}=e^{v(s) X^-}e^{v(s) X^+}$$
Notice that the matrix $\ Z(s)\in so(3)\ $ is normalized in such a way that 
the map $\ t\mapsto e^{t Z(s)}\in SO(3)\ $ represents a rotation around the 
axes  $R(s)=\vect{0}{\sin(\al)}{\cot (\frac{1}{2}v(s))}$ with angular 
velocity equal to one.
\el
Let us study the two possible cases described above:  
\medskip

($\clubsuit$) Suppose that the optimal trajectory starts with $u=-1$ (the case 
$u=1$ is symmetric) and has an even number $n$ of switchings. Then it must be 
\bqn
S=e^{s_f X^-}\underbrace{e^{v(s_i)X^+}\!\!\!\!\!\!\!\ldots\ldots\ e^{v(s_i)
X^+}}_{n-1\  times} e^{s_i X^-}N
\eqnl{pigreco}
where $N$ and $S$ denote respectively the north and the south pole, and we 
have that
$$e^{s_i X^-}S=e^{v(s_i)X^-}e^{v(s_i)X^+}\!\!\!\!\!\!\!\ldots\ldots\  
e^{v(s_i)X^+} e^{s_i X^-}N=$$ $$=e^{\frac{1}{2}n\th(s_i)Z(s_i)} e^{s_i X^-}N$$
from which we deduce that $s_i$ must satisfy 
$$\frac{1}{2}n\th(s_i)=\pi+2p\pi\ \mbox{ for some integer}\ p.$$
It is easy to see that a value of $s_i$ which satisfies previous equation 
with $p>0$ doesn't give rise to an optimal trajectory (since, roughly 
speaking, the corresponding number of switchings is larger than the number of 
switchings needed to cover the whole sphere). Therefore in previous equation
it must be $p=0$.\\
If $n$ is odd the relation \r{pigreco} becomes
\bqn
S=e^{s_f X^+}\underbrace{e^{v(s_i)X^-}\!\!\!\!\!\!\!\ldots\ldots\ e^{v(s_i)
X^+}}_{n-1\  times} e^{s_i X^-}N
\eqnl{pigreco2}
and, moreover, by simmetry:
$$N=e^{s_f X^-}e^{v(s_i)X^+}\!\!\!\!\!\!\!\ldots\ldots\  e^{v(s_i)X^-} 
e^{s_i X^+}S.$$
Then, combining with \r{pigreco2} and using the relation \r{belin}, we find:
$$N=e^{-s_i X^-}\underbrace{e^{v(s_i)X^-}\!\!\!\!\!\!\!\ldots\ldots\  
e^{v(s_i)X^+}}_{2n\  times} e^{s_i X^-}N=$$
$$=e^{-s_i X^-}e^{n\th(s_i)Z(s_i)} e^{s_i X^-}N .$$
Since $e^{s_i X^-}N$ is orthogonal to the rotation axis $R(s_i)$
corresponding to $Z(s_i)$, previous identity is satisfied if and only if 
$n\th(s_i)=2m\pi$ with $m$ positive integer. As in the previous case, for
an optimal trajectory, it must be $m=0$, and therefore the proof of
Proposition~\ref{p-1} is complete.
\medskip

($\spadesuit$) For simplicity call $s_i=s_f=s$. Assume, as before, that the 
optimal trajectory starts with $u=-1$ . If this trajectory has $n=2q+1$ 
switchings then it must be $$S=e^{s X^+}e^{q\th(s)Z(s)} e^{sX^-}N.$$
In particular the points $e^{-sX^+} S$ and 
$e^{sX^-} N$ must belong to a plane invariant with respect to
rotations generated by $Z(s)$ and therefore the difference
$e^{sX^-} N-e^{-sX^+} S$ must be orthogonal to the rotation axis $R(s)$.\\
Actually it is easy to see that this is true for every value $s\in[0,\pi]$, 
since both $e^{-sX^+} S$ and $e^{sX^-} N$ are orthogonal to $R(s)$.
Moreover, since the circle passing through $e^{sX^-} N$ and $e^{-sX^+} S$ 
corresponding to the rotations around $R(s)$ has radius 1, it is easy to 
compute the angle $\beta(s)$ between these points. In particular the distance 
between $e^{sX^-} N$ and $e^{-sX^+} S$ coincides with $2 \sin (\frac{\beta
(s)}{2})$ , and so one can easily obtain the expression:
$$\beta(s)=2 \arccos(\sin(\al)\cos(\al) (1-\cos(s))).$$
Then Proposition~\ref{p-2} is proved when $n$ is odd.\\
Suppose now that the optimal trajectory has $n=2q+2$ switchings, then we can 
assume without loss of generality that
$$S=e^{sX^-}e^{v(s)X^+}e^{q\th(s)Z(s)} e^{sX^-}N\ .$$
First of all it is possible to see that $e^{-v(s)X^+}e^{-sX^-}S$ is 
orthogonal to $R(s)$. 
So it remains to compute the angle $\tilde{\beta}(s)$ between the point 
$\,e^{sX^-}N\,$ and the point $\,e^{-v(s)X^+}e^{-sX^-}S\,$ on the plane 
orthogonal to $R(s)$. As before the distance between these points coincides 
with $2 \sin (\frac{\tilde{\beta} (s)}{2})$.\\ 
Instead of computing directly $\tilde{\beta}(s)$ we compute the difference 
between the angles $\tilde{\beta}(s)$ and the angle $\beta(s)$.
We know that
$$2 \sin (\frac{\tilde{\beta}(s)-\beta(s)}{2})=|e^{-v(s)X^+}e^{-sX^-}S-
e^{-sX^+} S|=$$ 
$$=|e^{-sX^-}S-e^{v(s)X^+}e^{-sX^+} S|=|e^{-sX^-}S-e^{s'(s)X^+}S|. $$
Using the fact that $s$ and $s'(s)$ satisfy the relation $v(s)=v(s'(s))$ one 
can easily find that $$|e^{-sX^-}S-e^{s'(s)X^+}S|=2\sqrt{1-\cos^2(\al)\sin^2 
\left(\frac{1}{2}v(s)\right)}.$$
Therefore 
$$\tilde{\beta}(s)=\beta(s)-2 \arccos{\left(\cos(\al)\sin\left(\frac{1}{2}
v(s)\right)\right)}.$$
This leads to $\beta(s)-\tilde{\beta}(s)=\th(s)/2$ and the 
proposition is proved also in the case $n$ is even.


\end{document}